\documentclass[12pt,preprint]{aastex}

\newcommand{\etal}{et~al.\/}

\newcommand{\dmt}[2]{\ensuremath{{#1}\degr\,{#2}\arcmin}}
\newcommand{\hmt}[2]{\ensuremath{{#1}^h\,{#2}^m}}

\newcommand{\sqdeg}{square degree}
\newcommand{\sqdegs}{square degrees}

\newcommand{\mjypbm}{{\rm mJy~beam}\ensuremath{^{-1}}}

\newcommand{\ie}{i.\,e.\,}

\newcommand{\uv}{\ensuremath{(u,v)}}

\newcommand{\sigatats}{\ensuremath{\sigma({\rm S_{ATATS}})}}
\newcommand{\mnatats}{\ensuremath{\overline{\rm S_{ATATS}}}}

\newcommand{\lognlogs}{log(N) -- log(S)}
\newcommand{\clognlogs}{Log(N) -- log(S)}
\newcommand{\postsize}{Postage stamps are $1\degr \times 1\degr$, centered on position in the first epoch where a source was detected. The integrated flux density of this source and its associated uncertainty (in mJy), and the number of sources within 75\arcsec\ of its position at this epoch, are shown above this panel. At each subsequent epoch the integrated flux densities and uncertainties (combined in quadrature) of the closest source or sources (if one or more exist within 75\arcsec) and the number of sources within that radius, are shown. For the NVSS panel, we consider only sources 10\,mJy or brighter. The ATA4 image is blank due to the loss of data from this epoch. Each panel is scaled independently using a linear scale from 1 -- 8 times the RMS flux density in that panel. The plot at upper left shows the transient lightcurve. The dashed line shows the sum of the integrated flux densities of all NVSS sources brighter than 10\,mJy within 75\arcsec\ of the ATA position. Epochs color-coded green (red) indicate ATA (non-)detections at that epoch.}
\newcommand{\paperi}{Paper I}

\shorttitle{The ATA 20-cm Survey II}
\shortauthors{Croft \etal}

\begin{document}
\title{The Allen Telescope Array Twenty-centimeter Survey -- A 700-Square-Degree, Multi-Epoch Radio Dataset -- II: Individual Epoch Transient Statistics }
\author{Steve Croft\altaffilmark{1}, Geoffrey C.\ Bower\altaffilmark{1},
 Garrett Keating\altaffilmark{1},
 Casey Law\altaffilmark{1},
 David Whysong\altaffilmark{1},
 Peter K.~G.~Williams\altaffilmark{1}, and
 Melvyn Wright\altaffilmark{1}}
\altaffiltext{1}{University of California, Berkeley, 601 Campbell Hall \#3411, Berkeley, CA 94720, USA }
\tabletypesize{\scriptsize}

\begin{abstract}

We present our second paper on the Allen Telescope Array Twenty-centimeter Survey (ATATS), a multi-epoch, $\sim 700$~\sqdeg\ radio image and catalog at 1.4\,GHz. The survey is designed to detect rare, bright transients as well as to commission the ATA's wide-field survey capabilities. ATATS explores the challenges of multi-epoch transient and variable source surveys in the domain of dynamic range limits and changing \uv\ coverage.

Here we present images made using data from the individual epochs, as well as a revised image combining data from all ATATS epochs. The combined image has RMS noise $\sigma = 3.96$\,\mjypbm, with a circular beam of 150\arcsec\ FWHM. The catalog, generated using a false detection rate algorithm, contains 4984 sources, and is $> 90$\%\ complete to 37.9\,mJy. The catalogs generated from snapshot images of the individual epochs contain between 1170 and 2019 sources over the 564~\sqdeg\ area in common to all epochs. The 90\%\ completeness limits of the single epoch catalogs range from 98.6 to 232\,mJy. 

We compare the catalog generated from the combined image to those from individual epochs, and from the NRAO VLA Sky Survey (NVSS), a legacy survey at the same frequency. We are able to place new constraints on the transient population: fewer than $6 \times 10^{-4}$ transients deg$^{-2}$, for transients brighter than 350\,mJy with characteristic timescales of minutes to days. This strongly rules out an astronomical origin for the $\sim 1$\,Jy sources reported by \citet{matsumura:09}, based on their stated rate of $3.1 \times 10^{-3}$\,deg$^{-2}$.

\end{abstract}


\section{Introduction}

In \citet{paperi}, hereafter \paperi, we introduced a 12-epoch survey undertaken with the Allen Telescope Array \citep[ATA;][]{welch}, the ATA Twenty-centimeter Survey (ATATS), and compared the image and catalog made with data from all 12 epochs to the NRAO VLA Sky Survey \citep[NVSS;][]{nvss}, a survey also at 1.4\,GHz (with 45\arcsec\ resolution, compared to the 150\arcsec\ resolution of our data) undertaken with the VLA between 1993 and 1997. Unfortunately both the raw and reduced data for epoch ATA4 were subsequently lost due to two disk failures, so for this paper we created a new image and catalog made with data from the remaining 11 epochs (which we term the {\em master mosaic} and {\em master catalog}).  We also generate images from data from individual epochs.  We compare the single-epoch catalogs to each other, and to NVSS, and search for transient and variable sources. 

The total integration time per pixel in the single-epoch mosaic images is approximately 2 minutes, and the epochs were separated by between 1 and 40 days (Table~\ref{tab:epochs}), with a median separation of 5 days. In comparing catalogs between epochs, therefore, we expect to be most sensitive to radio transients with timescales of minutes to days. Such events may have a range of progenitors \citep{lazio:09}.

Observing efficiency and telescope uptime have improved since the ATATS data were taken during commissioning of the telescope. Initial results from a larger survey using more recent data, the Pi GHz Sky Survey (PiGSS) are presented by \citet{pigss}.

\begin{deluxetable}{llllllll}
\tablewidth{0pt}
\tabletypesize{\scriptsize}
\tablecaption{\label{tab:epochs} ATATS epochs}
\tablehead {
\colhead{Epoch} & 
\colhead{UT Date} & 
\colhead{Pointings\tablenotemark{a}} & 
\colhead{$\sigma$\tablenotemark{b}} & 
\colhead{$N_{this}$\tablenotemark{c}} & 
\colhead{$N_{all}$\tablenotemark{d}} & 
\colhead{Completeness\tablenotemark{e}} &
\colhead{$S_{T}$\tablenotemark{f}}
\\
&
&
&
\colhead{(\mjypbm)} & 
&
&
\colhead{(mJy)} &
\colhead{(mJy)}
}
\startdata
ATA1  & 2009 Jan 12 & 251 & 9.69 & 1542 & 1170 & 163 & 181\\
ATA2  & 2009 Jan 19 & 268 & 6.69 & 1986 & 1415 & 174 & 247\\
ATA3  & 2009 Jan 26 & 250 & 7.18 & 2087 & 1648 & 143 & 350\\
ATA4  & 2009 Jan 31 & 0\tablenotemark{g} & \nodata & \nodata & \nodata & \nodata & \nodata \\
ATA5  & 2009 Feb 7  & 284 & 6.55 & 2555 & 1894 & 98.6 & 294\\
ATA6  & 2009 Feb 14 & 315 & 6.60 & 2638 & 1755 & 136 & 236\\
ATA7  & 2009 Feb 15 & 304 & 7.90 & 1914 & 1338 & 232 & 319\\
ATA8  & 2009 Mar 27 & 257 & 6.19 & 2656 & 2019 & 100 & 792\\
ATA9  & 2009 Mar 28 & 246 & 6.38 & 2380 & 1926 & 102 & 239\\
ATA10  & 2009 Mar 29 & 265 & 6.58 & 2285 & 1733 & 115 & 330\\
ATA11 & 2009 Mar 31 & 248 & 8.08 & 2090 & 1649 & 125 & 292\\
ATA12 & 2009 Apr 3  & 246 & 7.83 & 2306 & 1845 & 138 &271 \\
Master\tablenotemark{h} & \nodata  & 258 & 3.96 & 4984 & 4118 & 37.9 & 171\\
\enddata
\tablenotetext{a}{Number of mosaic pointings with rms noise $< 20$\,\mjypbm}
\tablenotetext{b}{RMS noise over the central quarter of the residual mosaic}
\tablenotetext{c}{Number of sources in the region of the mosaic which is good at this epoch}
\tablenotetext{d}{Number of sources in the 564~\sqdeg\ region of the mosaic which is good at all epochs and the master mosaic (Fig.~\ref{fig:coverage})}
\tablenotetext{e}{90\%\ completeness level, defined as the flux density above which 90\%\ of NVSS sources have a match in this ATATS epoch (\S~\ref{sec:cs})}
\tablenotetext{f}{Flux density of the brightest single-epoch transient candidate in this epoch (\S~\ref{sec:robust})}
\tablenotetext{g}{Data from this epoch were lost due to the failure of two hard drives (see text) and are not included in our analysis}
\tablenotetext{h}{Image made by combining data from all 11 epochs with data}
\end{deluxetable}

Throughout this paper, we use J2000 coordinates.

\section{Data Acquisition and Reduction}

The layout of the ATATS pointings, the details of the correlator settings, and the reduction scheme are detailed in \paperi, with some small differences as detailed below. 

The re-reductions were undertaken using a newer version of the RAPID \citep[Rapid Automated Processing and Imaging of Data;][]{keating:aas} software which provides for even better rejection of radio frequency interference (RFI) and other corrupted data than the previous versions due to a number of minor tweaks to the algorithms. This led to minor cosmetic improvements in image quality compared to \paperi. We measured $\sigma = 3.96$\,\mjypbm\ over the central quarter of a mosaic made from the residual images associated with the master mosaic. This is comparable to the  $\sigma = 3.94$\,\mjypbm\ from \paperi, despite the fact that the earlier paper contained an additional epoch of data in the master mosaic.

ATATS observations were spread over 12 epochs from 2009 January -- April (Table~\ref{tab:epochs}). At each epoch we aimed to observe as many of the 361 ATATS pointings as possible; in practice, at a given epoch, good data (RMS $< 20$\,\mjypbm\ in images of individual pointings --- a more stringent limit than the 50\,\mjypbm\ we used in Paper I) were obtained for between 246 and 315 pointings (Table~\ref{tab:epochs}). 

The size of the images made for each pointing was $512 \times 512$ pixels, where each pixel was $30\arcsec \times 30\arcsec$. 
All baselines shorter than 150 wavelengths were flagged a priori to remove any large-scale artifacts from the images (for example, due to solar interference). This typically resulted in the flagging of around 10\%\ of the visibilities. The longest baselines in the data are around 1400 wavelengths; see \paperi, Fig.~2 for plots of typical \uv\ coverage.

Not all pointings had good data at all epochs (for example, because of a failed observation, or severe corruption by RFI).\label{sec:corruptreg} The new version of RAPID produces better images in the regions surrounding 3C\,286 and 3C\,295, so in contrast to \paperi, they are included in the master mosaic here, which covers an area of 718~\sqdegs. However, these regions are affected by the presence of sidelobes, particularly in the single-epoch snapshot images, and were later excluded from our analysis (\S~\ref{sec:brightexc}).

\begin{figure}
\centering
\includegraphics[angle=270,width=\linewidth,draft=false]{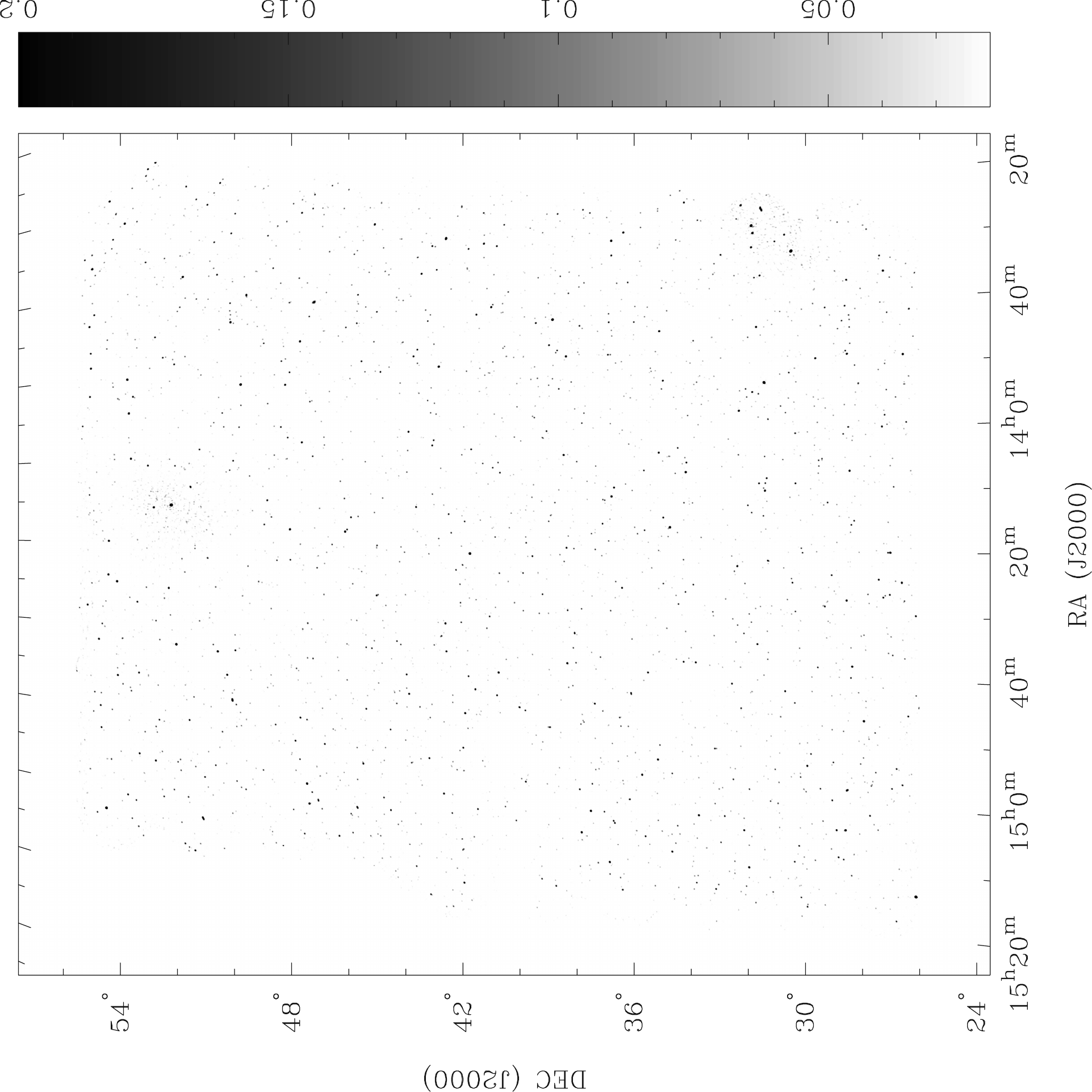}
\caption{\label{fig:deepfield}
The deep field image, made from a combination of all 11 epochs (compare Fig.~3 in \paperi). Note that regions with fewer than 8 epochs of good data were not included in the mosaic.
The greyscale runs from 20 to 200\,\mjypbm.
}
\end{figure}

We included in the master mosaic only pointings for which at least 8 out of 11 epochs were good (RMS $< 20$\,\mjypbm\ in the single-epoch image of that pointing). This led to a total of 258 pointings in the master mosaic. The majority of these (200) had good data for all 11 epochs. The master mosaic image is shown in Fig.~\ref{fig:deepfield}.

Mosaics of the individual epochs were made by combining all good images (RMS $< 20$\,\mjypbm) for that particular epoch, masking each outside a radius corresponding to the nominal half-power points (75\arcmin). 

As discussed in \paperi, the \uv\ coverage for individual snapshot observations is relatively poor compared to the \uv\ coverage obtained by combining data from all epochs. The mean RMS noise (computed over the central quarter of an image made by combining the residual images for each pointing)  for the individual epoch maps is 7.24\,\mjypbm\ (see Table~\ref{tab:epochs}). We would expect this to be $\sim \sqrt{11} = 3.3$ times higher than that for the master mosaic (3.96\,\mjypbm), \ie\ $\sim 13$\,\mjypbm, so we see that there must be some systematic effects limiting the sensitivity (presumably in the individual epoch images as well as the master mosaic). 

The MIRIAD task SFIND was run on the master mosaic and on the individual epoch mosaics. We used SFIND's false detection rate (FDR) algorithm \citep{sfindfdr} with a box size of 25 pixels, and the default 2\%\ acceptable percentage of false pixels (in contrast to the ``old'' algorithm used in \paperi, although we found little difference between catalogs produced using either method). Sources poorly fit by SFIND (indicated by asterisks in the SFIND output), typically comprising $\sim 3$\%\ of the total sources detected in each epoch, were rejected from the catalog. Also rejected from the final catalogs were the $\sim 5$\%\ of sources in regions where 10\%\ or more of the pixels in a $1\degr \times 1\degr$ box centered on the source were masked or blank, since these are typically regions with poor image quality close to the edges of the mosaic where noise peaks, corrected for primary beam attenuation, are mistakenly classified as real sources. The resulting number of sources, $N_{this}$, detected in each mosaic image (which depends on the area of that image), is shown in Table~\ref{tab:epochs}. 

The region with good ATATS data at all 11 epochs, and in the master mosaic, covers 635~\sqdegs. However,\label{sec:brightexc} regions surrounding 3C\,286 and 3C\,295, although they meet our criterion for good data, are in fact affected by the presence of sidelobes from these bright sources. When searching for transients (\S~\ref{sec:transients}), this results in spurious candidates in these regions, so we choose to exclude two regions of 4\degr\ in radius, centered on the positions of 3C\,286 and 3C\,295, leaving a 564~\sqdeg\ region\footnote{Although the excluded regions are 4\degr\ in radius, in practice less than $2 \times \pi \times 4^2$~\sqdegs\ was excluded from the map, because some parts of these regions were already considered bad, and the excluded regions extend beyond the edges of the mosaic.} which we use for our analysis. 

\begin{figure}
\centering
\includegraphics[angle=270,width=\linewidth,draft=false]{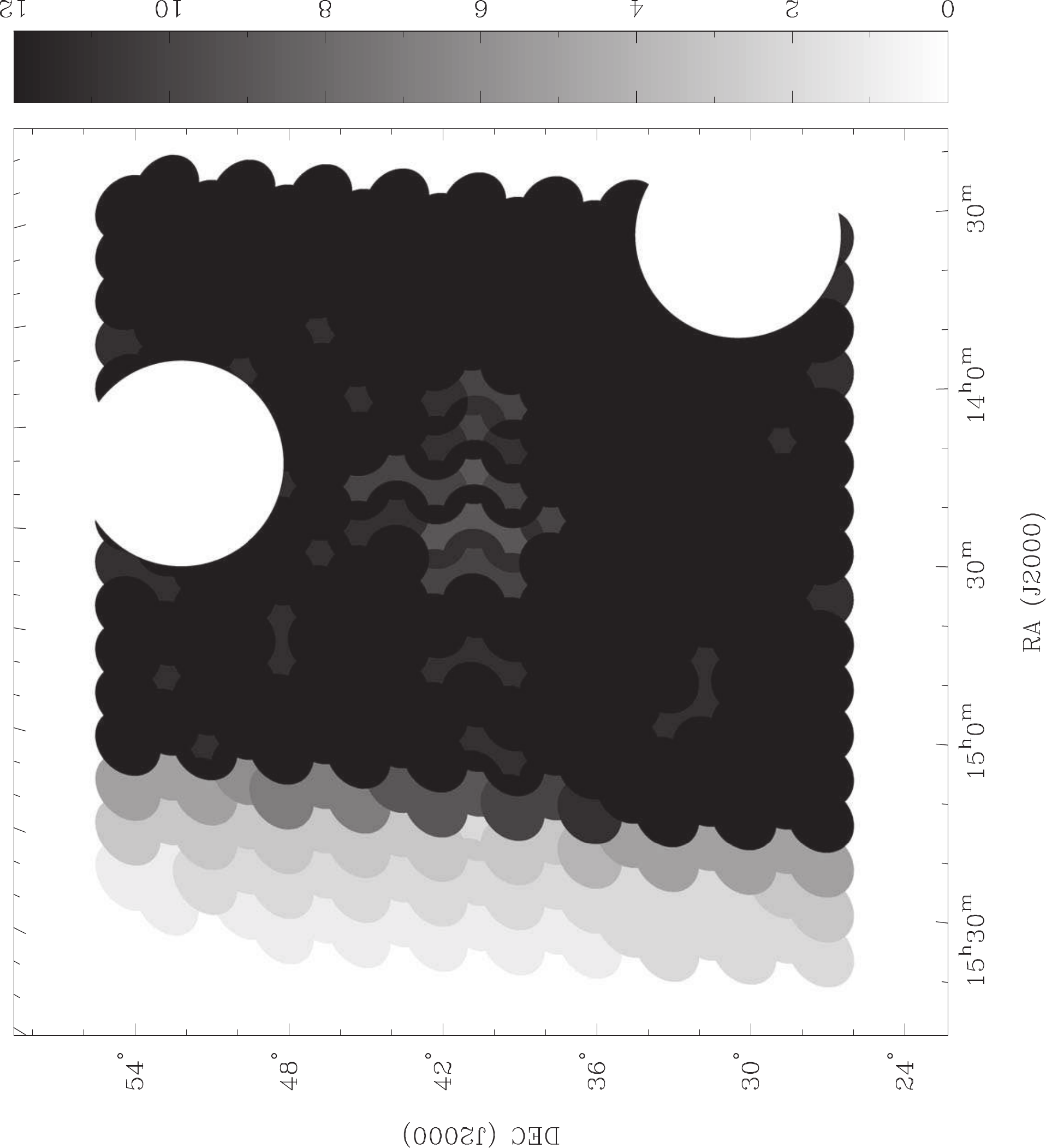}
\caption{\label{fig:coverage}
The region of sky with good data in ATATS. The greyscale shows the number of mosaic images with data having noise $\sigma \leq 20$\,\mjypbm\ as a function of sky position. The black area has good data in all 12 images (the 11 single epoch images and the master mosaic), and covers 564~\sqdegs. The two circular regions excised from the map are each 4\degr\ in radius, and are centered on 3C\,286 and 3C\,295.
}
\end{figure}

We also produced ``culled'' catalogs for all ATATS epochs and for NVSS, where we removed sources from the catalogs which were outside this region. The number of sources, $N_{all}$ in the 564~\sqdeg\ region with good data is shown in Table~\ref{tab:epochs}, and the sky coverage of the region itself is shown in Fig.~\ref{fig:coverage}. 

There are variations in quality of the mosaics from epoch to epoch, partly due to differing amounts of RFI and differing fractions of the data being affected by calibration problems, and partly due to small differences in \uv\ coverage. The latter was caused by some of the antennas being down for maintenance or otherwise unavailable due to array commissioning at some epochs, and changes in the sky position of our fields during the 81 days between the first and last epochs. Such quality variations result in different numbers of sources detected in each epoch, even though here the sky area remains the same. $N_{all}$ tends to decrease with increasing RMS $\sigma$, although the wide range in $N_{all}$ compared to that in $\sigma$ suggests that systematic effects are limiting the number of sources detected in epochs with higher $\sigma$. In our analysis, we use these culled catalogs so that we are comparing source counts and searching for transients in the same 564~\sqdeg\ region of sky where we have good sensitivity in all of our epochs. 

\subsection{Catalog matching and completeness}

\begin{figure}
\centering
\includegraphics[width=0.5\linewidth,draft=false]{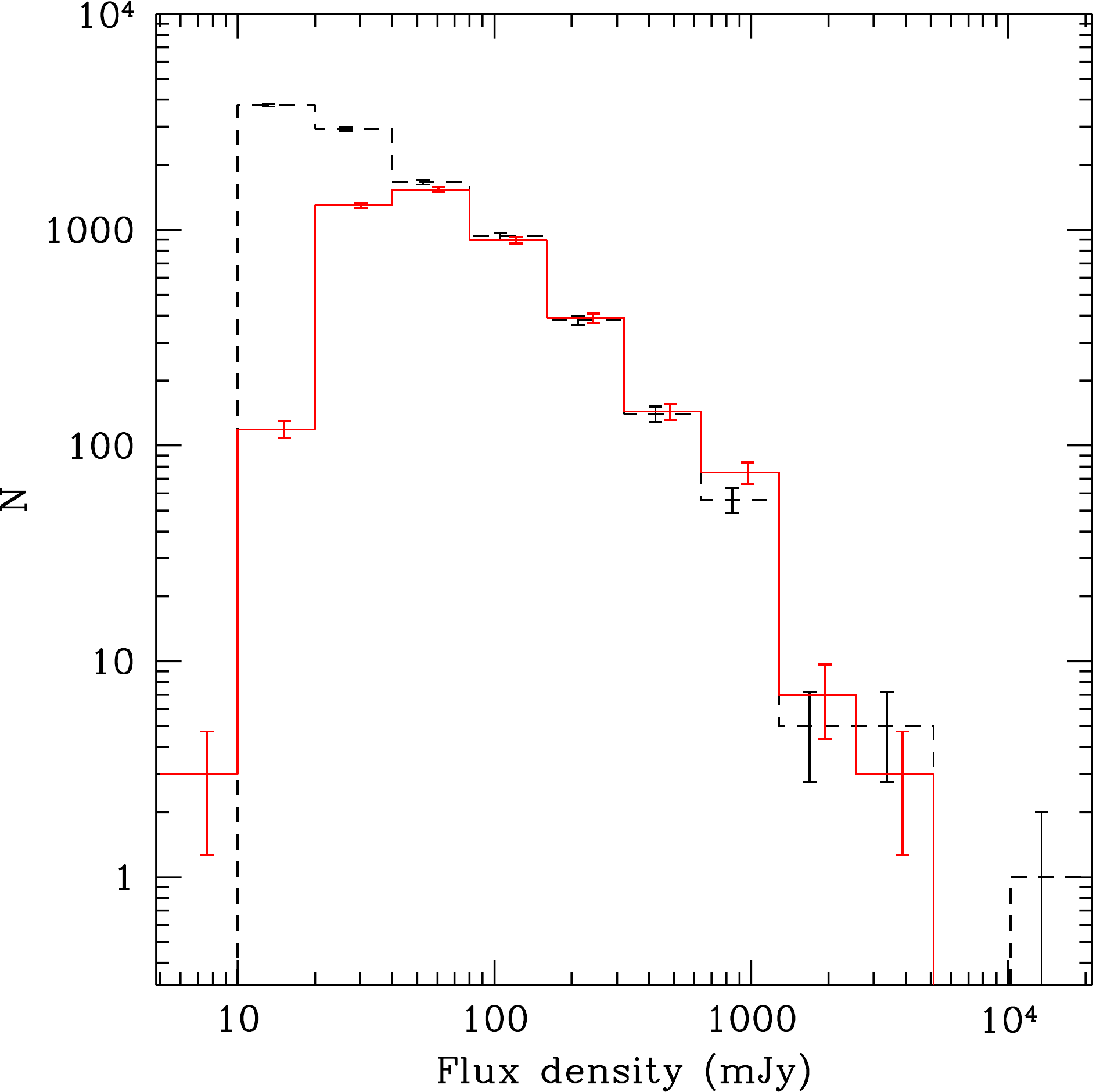}%
\includegraphics[width=0.5\linewidth,draft=false]{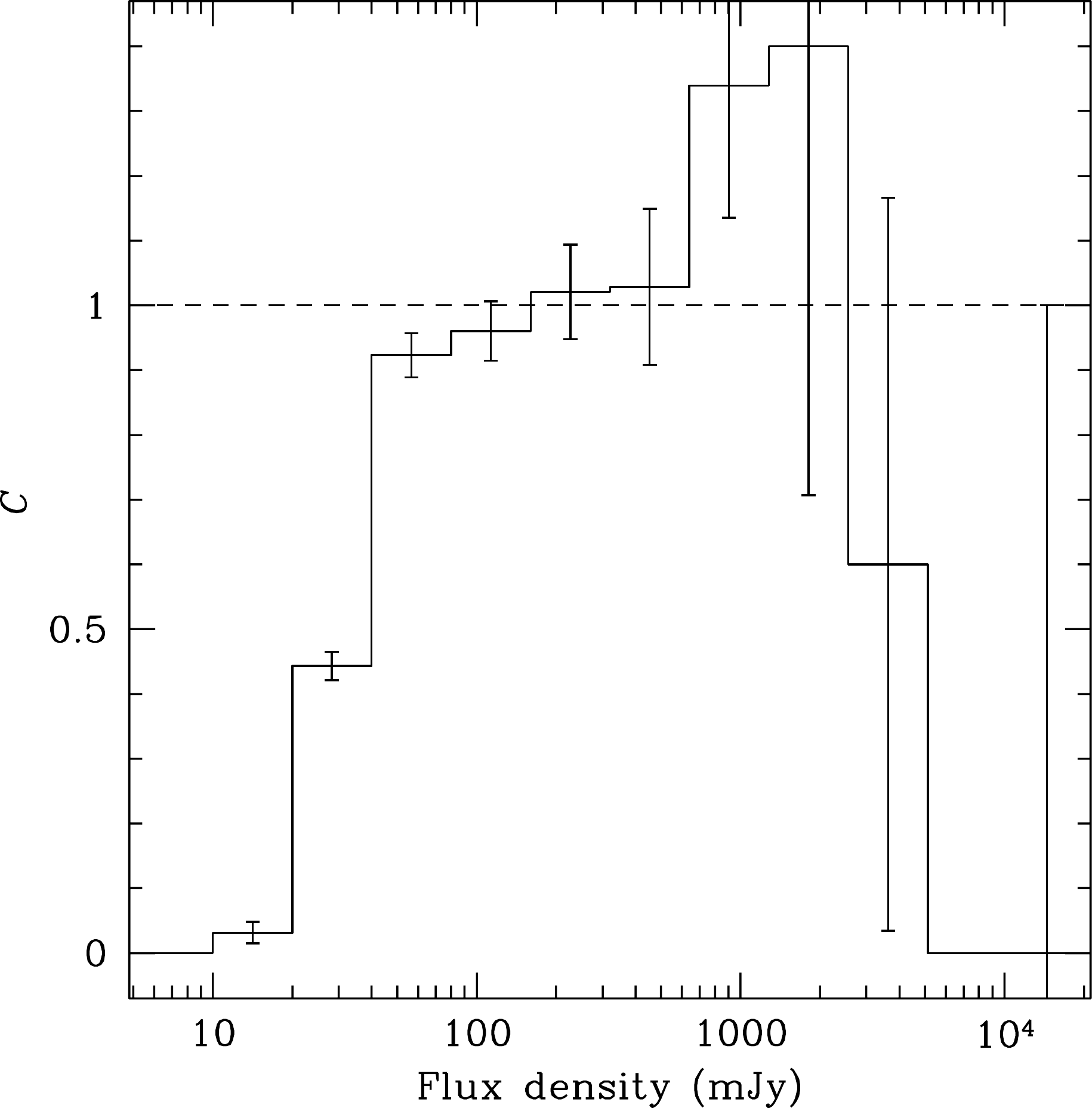}
\caption{\label{fig:compall}
{\em Left:} \clognlogs\ histogram of NVSS sources (black dashed line) in good regions (at all epochs) of the ATATS field, and sources detected in the ATATS master mosaic (red solid line) over the same area. Error bars assume Poisson statistics. Each bin covers a factor of 2 in integrated flux density, or 0.301 dex. The coarser resolution of the ATA tends to combine the flux from a tight cluster of NVSS sources into a single ATATS source. Combined with the ATA's additional sensitivity to structures larger than the NVSS synthesized beam, this tends to shift some sources from fainter into brighter bins. {\em Right:} \label{fig:complete}Ratio of the histograms for the ATATS master and NVSS catalogs, $C = N_{\rm ATATS} / N_{\rm NVSS}$ . This represents a measure of the completeness of the ATATS master catalog. Error bars are computed by propagating the Poisson errors from the ATATS and NVSS source count histograms. The tendency of the ATA to shift sources into higher flux bins results in some bins with $C > 1$. Compare \paperi, Fig.~10.
}
\end{figure}

As in \paperi, we can assess the completeness and reliability of ATATS by plotting the source count histograms of the ATATS and culled NVSS catalogs in Fig.~\ref{fig:compall}. The two histograms are consistent, within the errors, until the ATATS counts turn over between 40 and 80\,mJy, as previously determined in \paperi\ --- \ie, the results are consistent even though here we are using 11 epochs rather than 12, a newer version of the RAPID reduction software, the FDR algorithm in SFIND, and we are discarding sources in regions which have bad (or no) data in any given ATATS epoch rather than simply in bad regions of the master mosaic. The NVSS histogram truncates abruptly at 10\,mJy due to the cut we impose on the catalog. 

The ratio of the ATATS to NVSS histograms can also be interpreted as the efficiency with which we would detect transient sources of a given brightness, or as a measure of the survey completeness. We plot the ratio of the two histograms, which we denote $C$, in Fig.~\ref{fig:complete}. 

\begin{figure}
\centering
\includegraphics[width=\linewidth,draft=false]{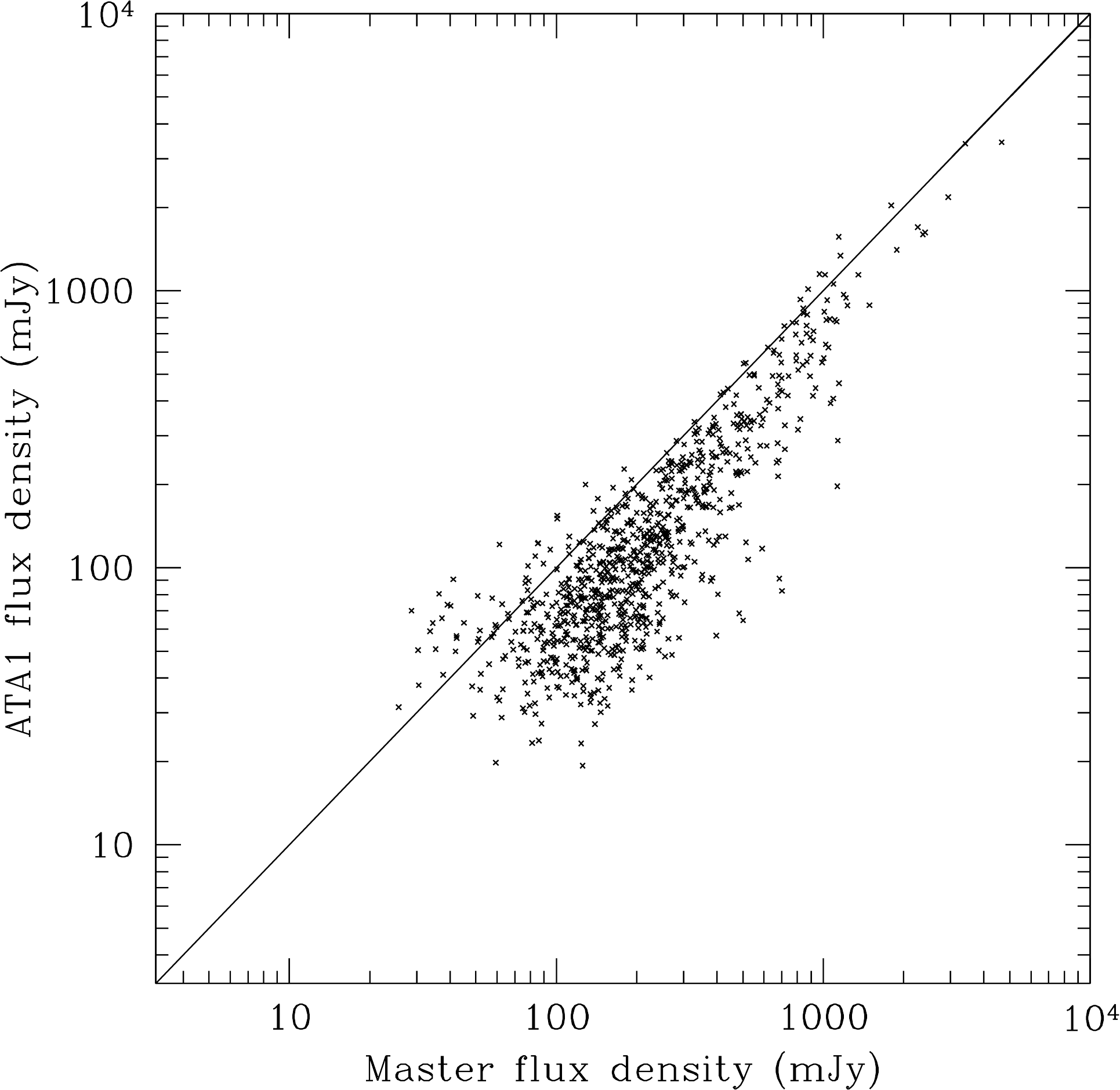}%
\caption{\label{fig:flux0111}
Comparison of the flux densities of all sources in the master catalog with the sum of the flux densities of all sources in the ATA1 catalog within a radius of 75\arcsec. Poor \uv\ coverage results in the single-epoch ATATS catalog underestimating the true flux densities of most sources.
}
\end{figure}

\begin{figure}
\centering
\includegraphics[width=\linewidth,draft=false]{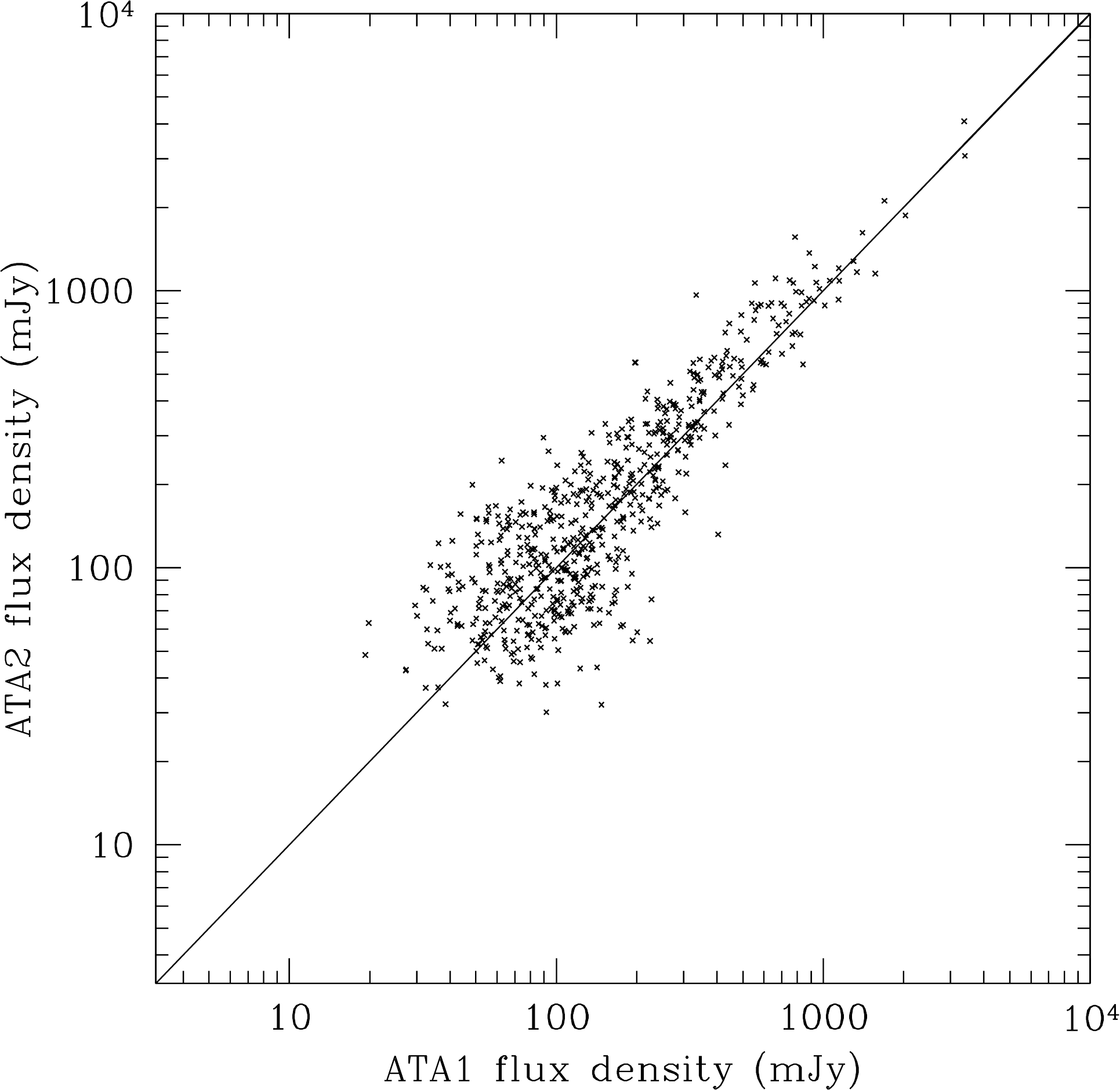}%
\caption{\label{fig:flux12}
Comparison of the flux densities of all sources in the ATA1 with the sum of the flux densities of all sources in the ATA2 catalog within a radius of 75\arcsec. There is less of a systematic bias of flux densities between epochs than when comparing single epoch maps to the master catalog as in Fig.~\ref{fig:flux0111}.}
\end{figure}

We can also assess completeness and the accuracy of our flux density measurements by comparing the catalogs source by source. We matched the catalogs to those from NVSS, using a circular 75\arcsec\ match radius and considering only NVSS sources brighter than 10\,mJy. We summed the flux densities of all NVSS sources within the match radius to help account for the difference in resolution between the two surveys. As in \paperi\ we obtained good agreement (negligible systematic flux density biases and acceptable random errors) between the master mosaic catalog and NVSS. Unfortunately the agreement between NVSS and the single-epoch catalogs, and between the single-epoch catalogs and the master catalog, is much poorer. Fig.~\ref{fig:flux0111} shows the cataloged flux densities of sources from a representative ATATS epoch (here, ATA1) compared with the sum of the flux densities of sources from the master catalog within 75\arcsec. It is clear that, especially at low flux densities, the single-epoch fluxes are biased low relative to the master mosaic fluxes by around a factor 2. Since we are able to recover unbiased flux densities relative to NVSS by combining the \uv\ data from all 11 ATA epochs, we conclude that the bias seen in the single epoch maps is likely due to a combination of poor snapshot \uv\ coverage, and CLEAN bias. The individual epoch catalogs are reasonably consistent with each other (Fig.~\ref{fig:flux12}), with some scatter.

We attempted to improve the image fidelity by performing a joint deconvolution of all pointings. We also tried a joint deconvolution of each pointing and its immediate neighbors, and then generated a linear mosaic of the central pointings of each of these images (in an effort to better control sidelobes). The flux densities were somewhat less biased relative to NVSS and the master mosaic, but in general the image quality was worse, with more scatter in the flux densities, and artifacts affecting more of the images. We are able to obtain higher quality images from the ongoing PiGSS survey, which has better \uv\ coverage (three snapshot observations of each field per epoch rather than one), and more baselines (due to ongoing array commissioning activities). We hope to apply techniques to PiGSS such as ``peeling'' off bright sources from the \uv\ data, but we judged such efforts beyond the scope of our investigations of the pilot ATATS study.

If we use the source count histogram method described above with these data to estimate completeness, the bias in flux densities towards lower values will tend to shift the \lognlogs\ histogram for ATATS in plots like Fig.~\ref{fig:compall} to the left and result in us obtaining values for the completeness limit that are erroneously high. Even though we are unable to accurately determine flux densities in the single-epoch maps, however, we can still simply examine whether or not there is a match to each NVSS source in the culled catalog as a function of cataloged NVSS flux density. Instead of plotting completeness, $C = N_{\rm ATATS} / N_{\rm NVSS}$, bin by bin as in Fig.~\ref{fig:compall}, we define a new completeness estimator, 

\begin{equation}
C(S) = 1 - \frac{N_{\rm u}(S)}{N_{\rm a}(S)}
\end{equation}

where $N_{\rm u}(S)$ is the number of NVSS sources in the culled catalog with NVSS flux density brighter than $S$, and no match in the ATATS catalog within 75\arcsec; and $N_{\rm a}(S)$ is the total number of NVSS sources in the culled catalog with NVSS flux density brighter than $S$. This statistic does not rely on an accurate measure of the flux densities from ATATS; it simply determines whether or not there is a match in ATATS to a particular NVSS source, and what fraction of sources brighter than this are not missing from the ATATS catalog.

\begin{figure}
\centering
\includegraphics[width=0.5\linewidth,draft=false]{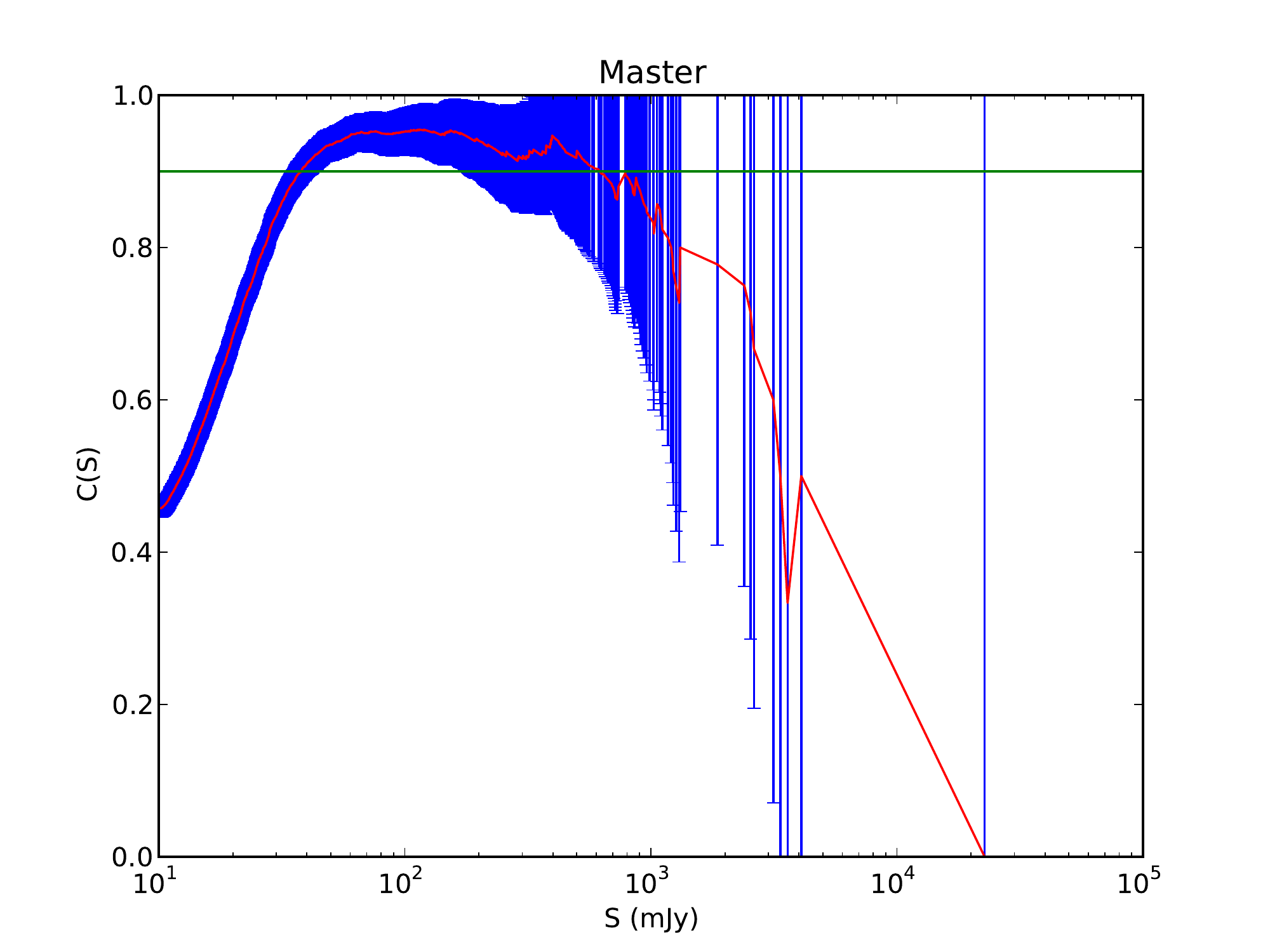}
\includegraphics[width=0.5\linewidth,draft=false]{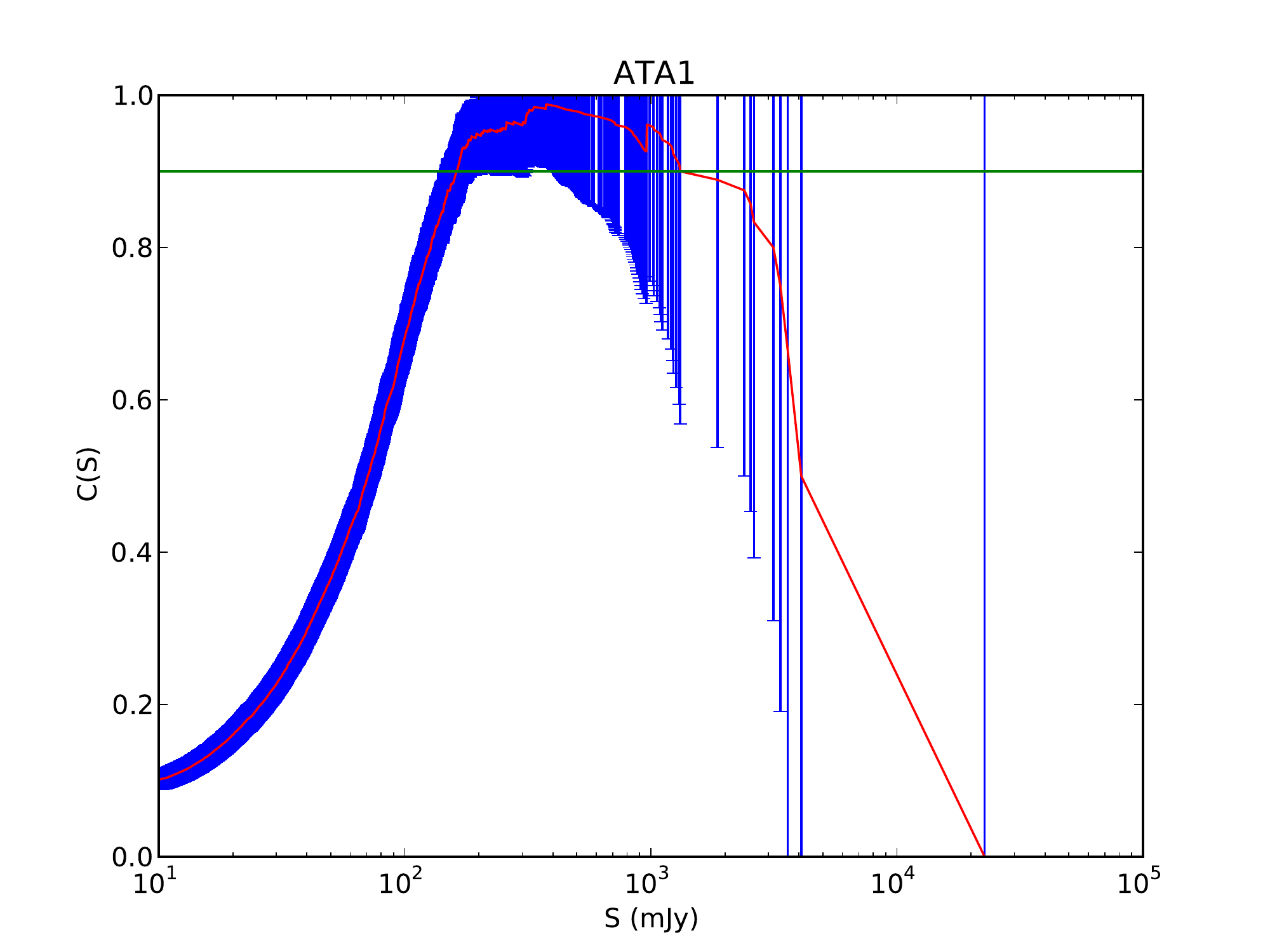}%
\includegraphics[width=0.5\linewidth,draft=false]{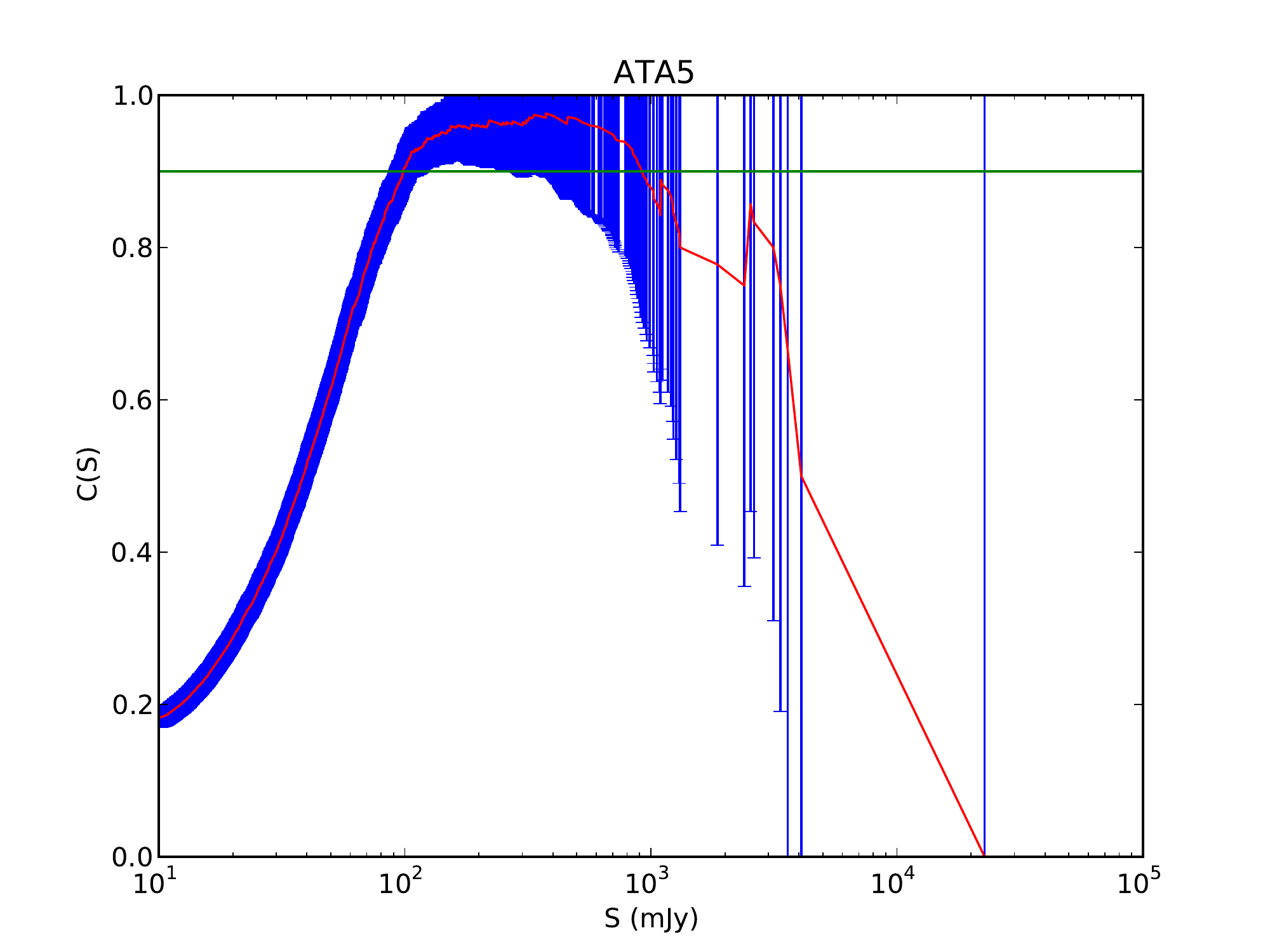}
\caption{\label{fig:csmaster}
\label{fig:csata1}
\label{fig:csata5}
Completeness, $C(S) = 1 - (N_{\rm u}(S) / N_{\rm a}(S))$, \ie, the fraction of NVSS sources brighter than flux density $S$ with a match in ATATS. Shown here are the Master, ATA1 and ATA5 catalogs. The horizontal line shows 90\%\ completeness, and the error bars assume Poisson statistics. The data points have not been binned.
}
\end{figure}

\label{sec:cs}We plot $C(S)$ for the master catalog in Fig.~\ref{fig:csmaster}. The culled NVSS catalog is sorted in increasing flux density order, and the fraction of NVSS sources brighter than that flux density with ATATS matches is plotted. At first, completeness increases with increasing NVSS flux density, until it crosses the 90\%\ threshold at 37.9\,mJy (comparable to the value of 40\,mJy we previously obtained using other methods). After a broad plateau at around 100\,mJy, completeness starts to fall off again. This is because NVSS sometimes resolves bright sources into multiple components which are outside our 75\arcsec\ match radius. Also, map quality tends to be poorer in regions surrounding very bright sources (with higher RMS and worse sidelobes), and such sources are more likely to suffer from poor fits and failed deblending. This causes completeness to fall off because these sources are not present in the culled SFIND catalogs despite being clearly visible in the ATATS images.

We also plot $C(S)$ for the ATA1 catalog in Fig.~\ref{fig:csata1}. The 90\%\ completeness limit, 162\,mJy, is much higher than for the master catalog, which we attribute to the poorer \uv\ coverage of the single-epoch maps. We repeat the completeness measurement for all 11 ATATS epochs, and tabulate the measured 90\%\ completeness values in Table~\ref{tab:epochs}. The best single epoch completeness is for epoch ATA5, and we also plot $C(S)$ for this epoch in Fig.~\ref{fig:csata5}.

\subsection{Multi-epoch matches}

\begin{figure}
\centering 
\includegraphics[width=0.23\linewidth]{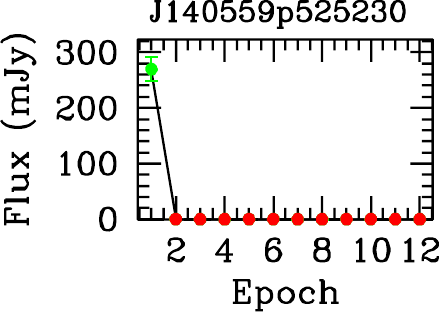}%
\includegraphics[width=0.77\linewidth]{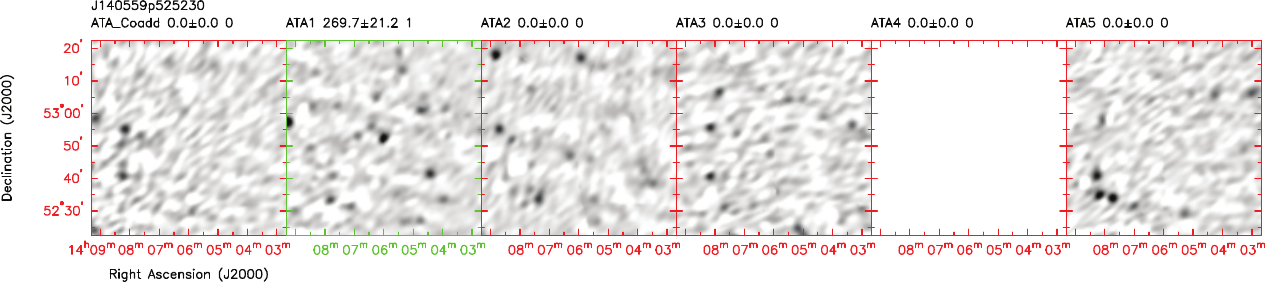}
\hspace*{0.23\linewidth}%
\includegraphics[width=0.77\linewidth]{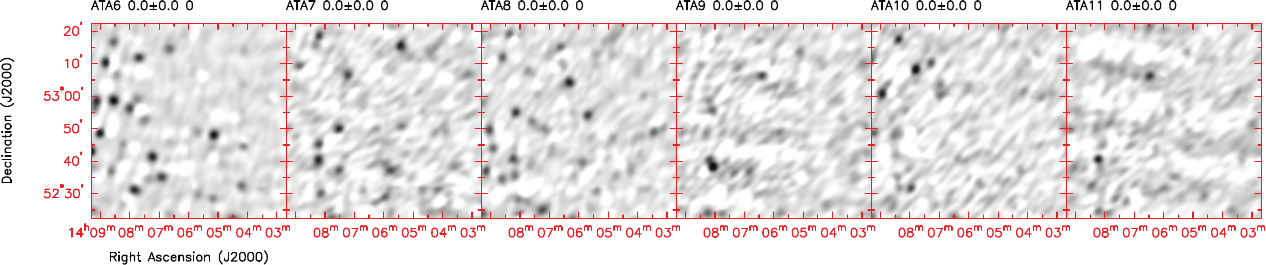}
\hspace*{0.23\linewidth}%
\includegraphics[width=0.77\linewidth]{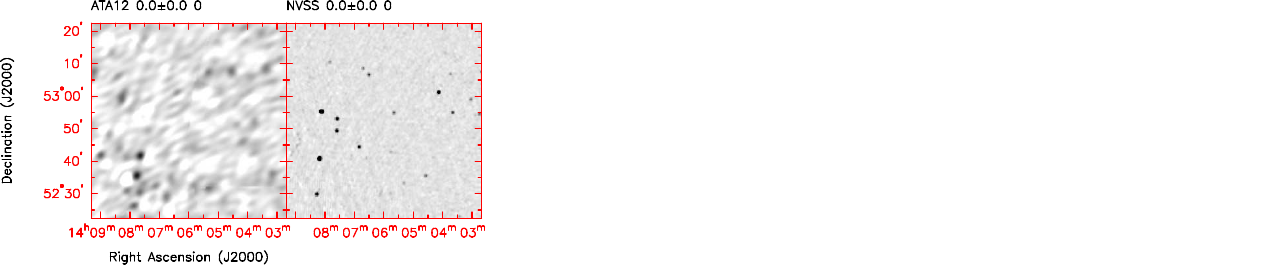}
\caption{\label{fig:onetranscandbad}
A single-epoch transient candidate (ATA flux density brighter than 232\,mJy in one epoch, and no match at other epochs or in NVSS). The changing pattern of sources in the field illustrates that this region (close to 3C\,295) is affected by sidelobes from that 23\,Jy source. This example is from one of the regions worst affected by sidelobes; compare Fig.~\ref{fig:bigvar}.  \postsize
}
\end{figure}

We created a software package, SLOW (Source Locator and Outburst Watcher), which although not optimized for speed, is capable of generating light curves and postage stamp images for sources detected in ATATS, and corresponding NVSS detections (or non-detections). SLOW was used to produce an sqlite3 database containing the information from the SFIND catalogs, as well as multi-epoch light curves and postage stamp images from the ATATS data, such as those in Fig.~\ref{fig:onetranscandbad}.

SLOW parses the culled catalogs one by one, source by source. If it finds a source with an existing match in the database (the default is another source within 75\arcsec), the new source is assigned the same source designation as the existing source. If no such match is found, the source is assigned its own designation and the match algorithm continues to work its way through the catalogs and epochs until all sources have been examined. The database can then be queried to extract flux densities and other measured parameters for sources found in multiple epochs, as well as for sources that are absent in some epochs but present in others. In particular, we search for sources without a match in NVSS which are present in one or more ATA epochs.

\subsection{Variability}

\begin{figure}
\centering
\includegraphics[width=\linewidth]{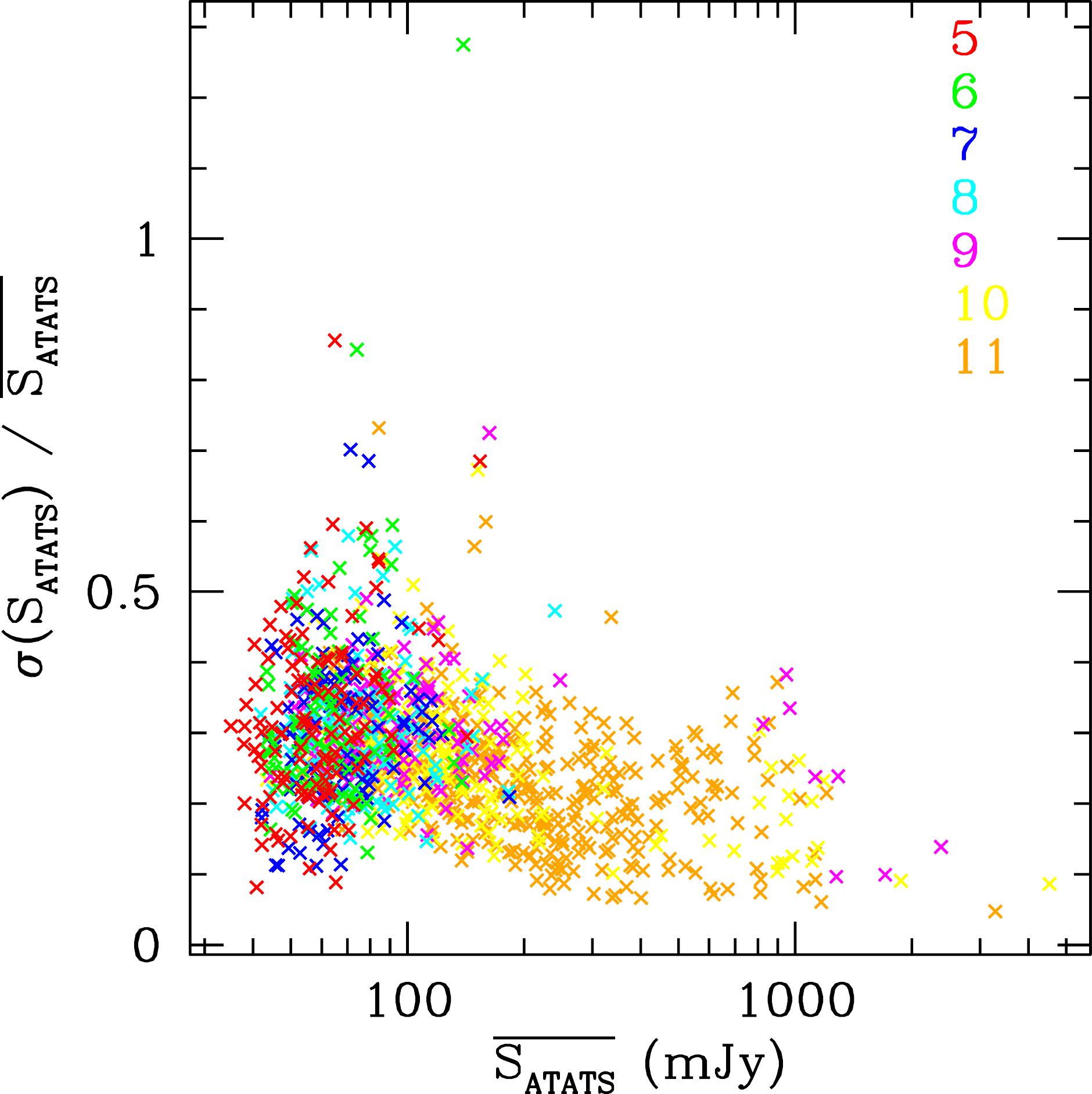}
\caption{\label{fig:slowstd}
The ratio of the standard deviation of flux densities of sources detected in ATATS, \sigatats, to the mean flux density of the sources, \mnatats, plotted as a function of \mnatats, for sources with detections in 5 or more ATATS epochs. Points are color coded according to the number of ATATS epochs in which they were detected.
}
\end{figure}

The problems with flux calibration in the individual epochs make placing constraints on epoch-to-epoch variability difficult. However, we can still attempt to identify highly variable sources. We query the sqlite3 database for sources with detections in at least 5 ATATS epochs (to avoid small number statistics for sources detected in 4 epochs or fewer) and found 1061 sources. For the 991 sources not within 4\degr\ of 3C\,286 or 3C\,295, we compute the mean flux density, \mnatats\ (ignoring epochs where the source was undetected) and its standard deviation, \sigatats. We divide \sigatats\ by \mnatats\ to get a measurement of the fractional variability of the sources in the 11 ATATS epochs, and plot this quantity as a function of \mnatats\ in Fig.~\ref{fig:slowstd}. This plot shows that sources brighter than $\sim 120$\,mJy tend to be detected in at least 9 of the 11 epochs. The completeness limits in Table~\ref{tab:epochs} suggest that there is a $\gtrsim 90$\%\ chance per epoch of seeing a 120\,mJy source in only 4 of the epochs, but these completeness limits are computed for all sources, including sources with complex morphologies. Sources which are well-detected in several epochs are those which are among the most likely to be detected in additional epochs. Fig.~\ref{fig:slowstd} also shows that bright sources tend to have lower fractional variability compared to fainter sources, as would be expected if measurement uncertainties and intrinsic variability do not increase linearly with mean flux density.

Sources detected in 8 epochs or fewer tend to be found at low flux densities in Fig.~\ref{fig:slowstd}. Such sources could have relatively constant measured flux in the epochs in which they are detected, but fall below the completeness limit in other epochs, in which case they will exhibit relatively small fractional variability. Some have relatively large fractional variability, which may be a cause of their being detected at fewer than 11 epochs (if they are intrinsically variable and drop below the detection threshold in some epochs) or an effect of measurement uncertainties which are large relative to the flux densities of such faint sources, even if their true flux densities are not intrinsically highly variable.

\begin{figure}
\centering
\includegraphics[width=0.23\linewidth]{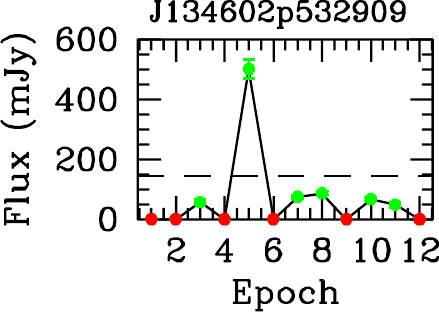}%
\includegraphics[width=0.77\linewidth]{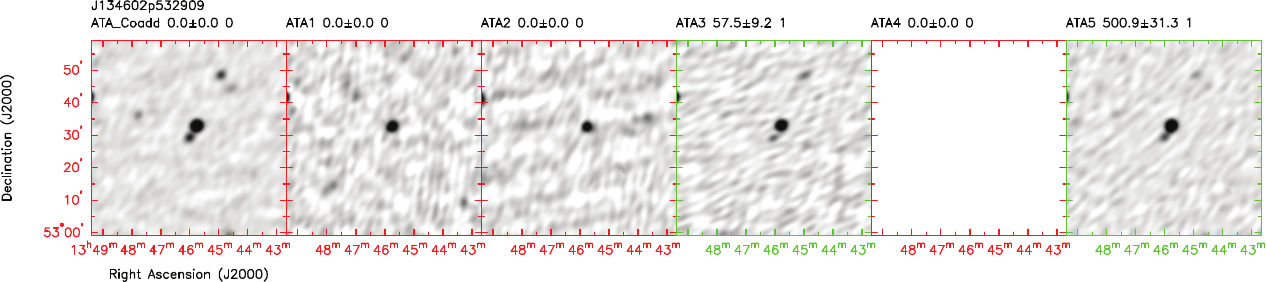}
\hspace*{0.23\linewidth}%
\includegraphics[width=0.77\linewidth]{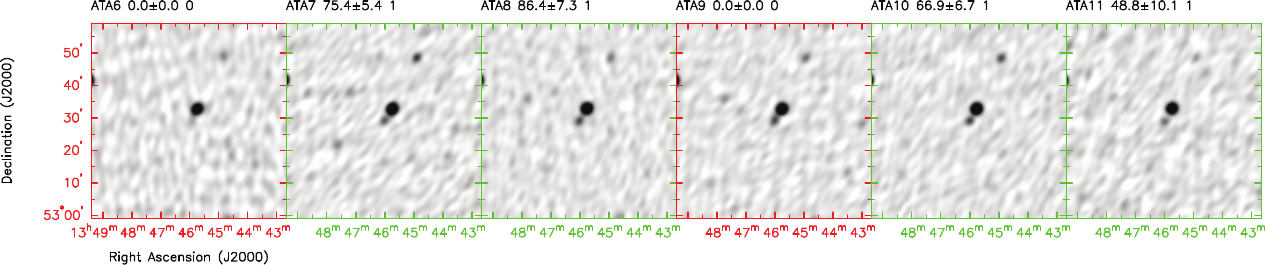}
\hspace*{0.23\linewidth}%
\includegraphics[width=0.77\linewidth]{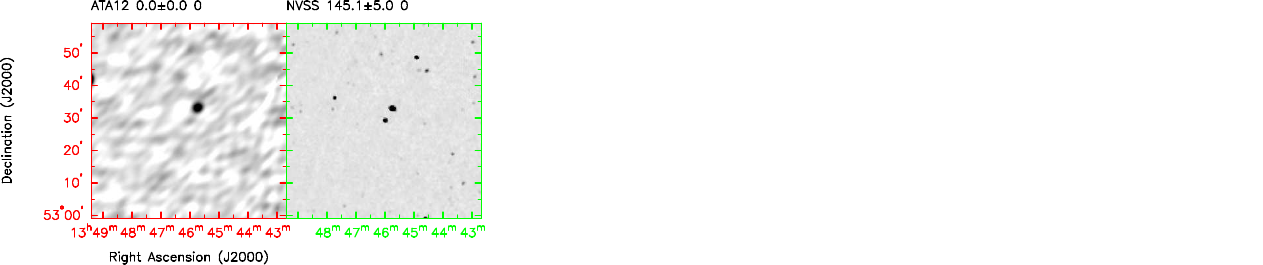}
\caption{\label{fig:bigvar}
A source with $\sigatats / \mnatats > 0.5$. Although apparently highly variable, the variability is likely due to the complex morphology and the fact that the source is sometimes not detected due to poor \uv\ coverage, or sometimes not properly deblended or fit due to the nearby bright source to the northwest. For example, in ATA5, the source in the center of the map has not brightened by a factor of $\sim 9$ relative to ATA3 -- it is merely poorly deblended from the source to the northwest in that epoch.
\postsize
}
\end{figure}

We examined postage stamps and light curves for the 8 sources with $\mnatats \geq 98.6$\,mJy (the best 90\%\ completeness limit from Table~\ref{tab:epochs}) and $\sigatats / \mnatats > 0.5$, \ie, the outliers in Fig.~\ref{fig:slowstd}. As in \paperi, we find that these candidate highly variable sources are in fact dominated by objects with close neighbors, such as that shown in Fig.~\ref{fig:bigvar}, or with complex morphologies. Such sources are sometimes not correctly fit or deblended by the source finding algorithm at some epochs, especially as the snapshot \uv\ coverage means that the relative fluxes and sizes of components are not always completely consistent from epoch to epoch.

\subsection{Transients\label{sec:transients}}

\begin{figure}
\centering
\includegraphics[width=\linewidth]{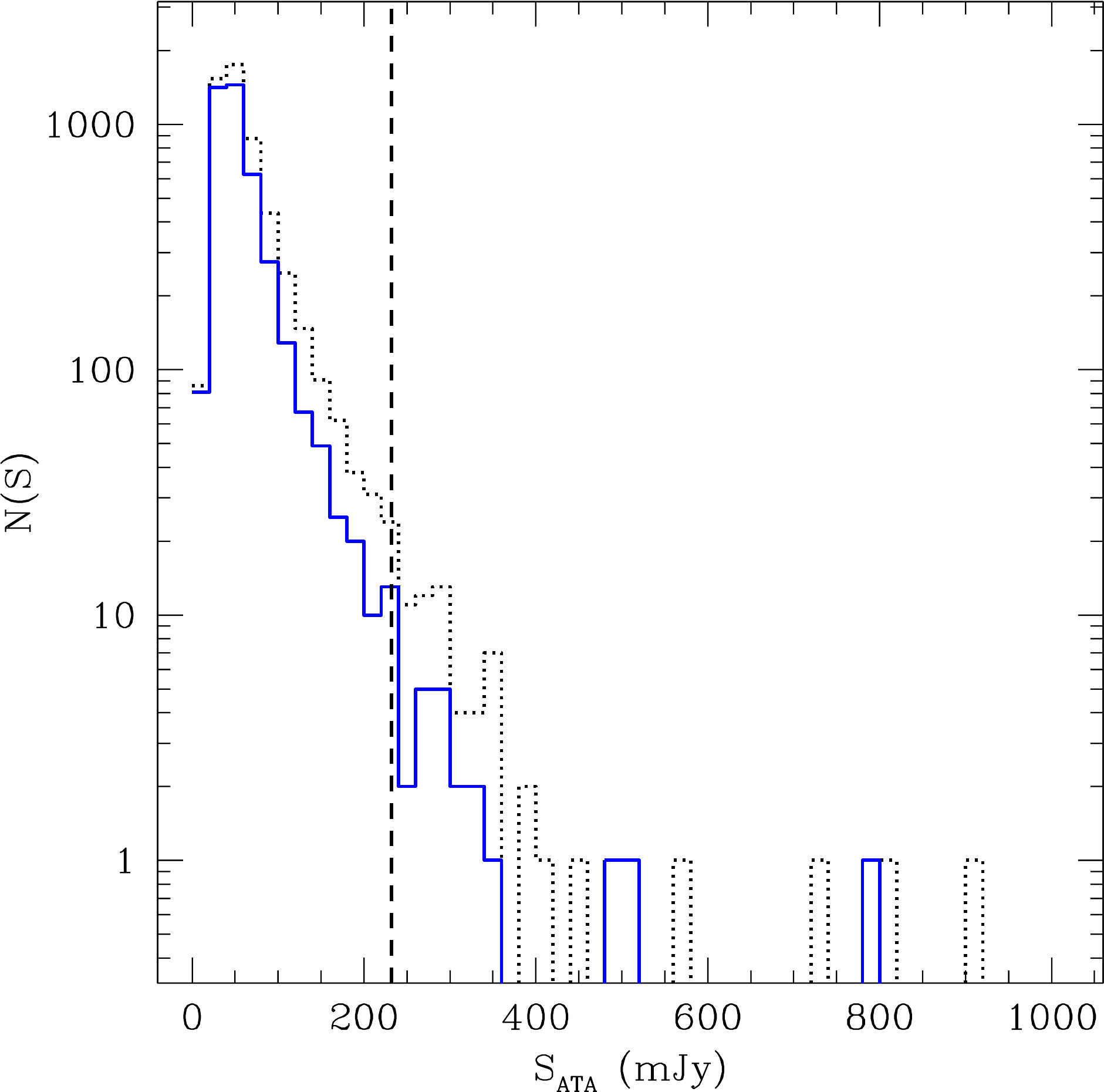}
\caption{\label{fig:transhist}
Histograms of the flux densities of sources detected only in a single ATATS epoch, and not in NVSS. The black dotted histogram shows sources detected over the 635~\sqdeg\ region with good data in all epochs, and the blue solid histogram shows sources detected in the 564~\sqdeg\ region which excludes regions close to 3C\,286 and 3C\,295. The vertical dashed line shows the 232\,mJy cut (corresponding to the completeness limit in ATA7 --- see Table~\ref{tab:epochs}) above which we examined postage stamps and lightcurves of transient candidates. None of these turn out to be astronomical transients.
}
\end{figure}

Due to the snapshot \uv\ coverage and its effect on our ability to obtain accurate flux densities, we must be conservative in using these data to place limits on transients. Astronomical transients with timescales shorter than a few days would be expected to be seen in a single ATATS epoch. In Fig.~\ref{fig:transhist}, we show histograms of the flux densities of sources detected only in a single ATATS epoch, and not in NVSS (to a flux density limit of 10\,mJy). There are large numbers of faint transient candidates, and while some of these may be astronomical events, it is also likely that many are spurious. Some intrinsically steady sources that are fainter than the single epoch completeness limits might be detected in only one or two ATATS epochs, and hence be considered transient candidates --- except for the fact that one would expect most such sources to have a match in the much more sensitive NVSS catalog. Rather, some of the faint ATATS transient candidates (with no NVSS match) might be true astronomical events, but we find that most are likely due to imaging artifacts caused by the snapshot \uv\ coverage. Fig.~\ref{fig:transhist} also shows, however, a handful of very bright sources, and we must examine these carefully to determine if any are astronomical events.

Our single epoch images are 90\%\ complete at NVSS flux densities of 232\,mJy or better (Table~\ref{tab:epochs}), so astronomical sources brighter than 232\,mJy which are not strongly variable would be expected to be seen in multiple epochs. Therefore, sources brighter than 232\,mJy appearing only in a single ATATS epoch are relatively good transient candidates.

\begin{figure}
\centering
\includegraphics[width=\linewidth]{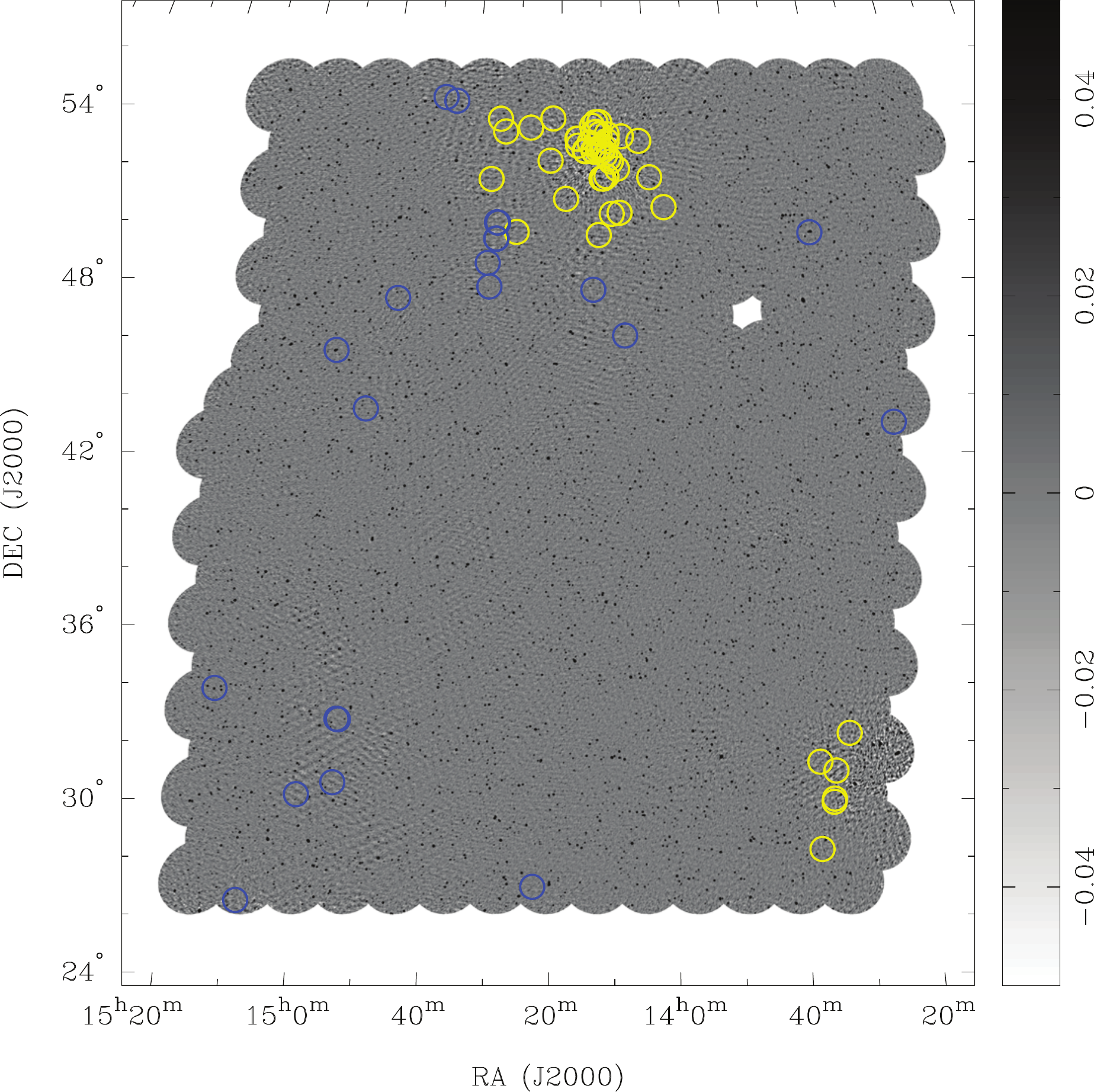}
\caption{\label{fig:onetrans}
The master mosaic image, with grayscale running from -50 to 50\,\mjypbm\ in order to highlight the small imperfections in image quality, and sidelobes surrounding bright sources. Overplotted as circles are the 70 single-epoch transient candidates. Yellow circles are candidates within 4\degr\ of the catalog positions of 3C\,286 or 3C\,295, which are very likely to be sidelobes of these sources. The remainder of the candidates are plotted as blue circles. Note that the candidates were detected in images from the individual ATATS epochs, and are shown here overplotted on the master mosaic simply for clarity. It is likely that all of these candidates shown are sidelobes from bright sources, imaging artifacts, or failed deblending of sources with complex morphologies.}
\end{figure}

 We queried the SLOW database for sources which appear in a single ATATS epoch with flux density brighter than 232\,mJy but no match at other epochs or in NVSS to 10\,mJy. This results in 70 candidates, and we examined the postage stamps and light curves for these, as well as the wider regions surrounding the candidates on the mosaic images. We plot the locations of these candidates on the master mosaic (note, though, that since these candidates show up in a single epoch, they are not necessarily visible, or in regions badly affected by sidelobes, in the master mosaic itself) in Fig.~\ref{fig:onetrans}. As can be seen, the candidates cluster in regions with high sidelobes, particularly around the bright sources 3C\,286 (\hmt{13}{31}, \dmt{30}{31}) and 3C\,295 (\hmt{14}{11}, \dmt{52}{12}).  We plot the 6 sources within a 4\degr\ radius of 3C\,286, and the 42 sources within a 4\degr\ radius of 3C\,295, in yellow, to denote that they are almost certainly sidelobes of these sources. However, we carefully examined postage stamps of these 48 candidates, as well as the 22 candidates further away from these very bright sources. We confirmed that most appeared to be due to sidelobes (for an example, see Fig.~\ref{fig:onetranscandbad}) --- often, the sources do not appear pointlike, and additionally in many of these postage stamps there are obvious imaging problems, including changing patterns of sources due to other sidelobes. The remainder of the cases all either appear to suffer from imaging artifacts, or are similar to the source in Fig.~\ref{fig:bigvar}, except in these cases the poor fitting or failed deblending results in a spurious single-epoch transient candidate rather than a candidate highly-variable source.

To search for transients on timescales comparable to the spacing between epochs, we also queried the SLOW database for sources which are detected in two or three ATATS epochs, with flux density brighter than 232\,mJy in at least one of those epochs, and no match in NVSS down to 10\,mJy or in the remaining ATATS epochs. Four sources were detected in two ATATS epochs only, but all were less than 4\degr\ away from 3C\,295. Examination of postage stamps for these candidates confirms that they are almost certainly sidelobes of 3C\,295. Just one candidate present in three epochs was found (a 247\,mJy detection in one epoch, and fainter matches in one other single epoch and in the master mosaic) --- this source is 0.8\degr\ from 3C\,295, and again we reject this as a likely sidelobe. We also searched for sources with maximum ATATS flux at least 232\,mJy seen in four ATATS epochs and not in NVSS, but none were found.

Sources detected in five or more ATATS epochs but not in NVSS, whatever their flux density, are good candidates for transients with timescales longer than the few days between ATATS epochs, but shorter than the $\sim 15$~years between the NVSS and ATATS observations. It is unlikely that imaging artifacts or sidelobes would recur at the same position in five or more independent epochs, so these sources are likely astronomical. Since the NVSS flux density limit is much fainter than the completeness limits for ATATS, if such sources truly have no match in NVSS, they are likely to be transients.

We queried the SLOW database for sources with five or more ATATS detections and no NVSS match. A single source was found with a detection in 5 epochs but not in the other epochs or in NVSS. Also one was found 10 times in ATATS (including in the master mosaic) but not in two other epochs or in NVSS, and one was found 12 times in ATATS, including the master mosaic. Examination of the postage stamp images showed all three of these cases to be sources with complex morphology, visible in all epochs including NVSS but with poor deblending at the epochs with no cataloged source within 75\arcsec. No other sources were seen with cataloged detections at any flux density in 5 -- 12 ATATS epochs but no NVSS match. We conclude that ATATS detected no transients with timescales longer than a few days.

These results are consistent with our results from \paperi\ (where we also considered sources in the original master mosaic as faint as 120\,mJy) that ATATS contains no bright transients. If a source within the good region of our map attained a mean flux density of at least 232\,mJy over the one minute snapshot integration time of the ATATS pointings, we should have seen it, at 90\%\ confidence. This confidence limit is conservative, because 10 of the 11 ATATS epochs had fainter completeness limits than 232\,mJy. Also, the 10\%\ of NVSS sources brighter than this limit without a match in the ATATS master catalog (consisting of 24 NVSS sources) are in fact all clearly visible in the ATATS master mosaic and all of the single epoch ATATS images. Most of the sources are extended or have complex morphologies, so the lack of a match in the ATATS catalog is due to failed deblending or poor fits. In other words, if a transient brighter than 232\,mJy with a simple Gaussian morphological profile occurred in one of the ATATS epochs, in a region of the mosaic where we have good sensitivity, we would be almost certain to see it, even though Fig.~\ref{fig:flux0111} suggests that we would underestimate its flux density.

If we consider sources which may have been transients in NVSS but are no longer seen in ATATS, as noted above, there are no true NVSS sources brighter than our worst single-epoch completeness limit, 232\,mJy, which are not present in any of the ATATS images. A more stringent limit was obtained in \paperi, however --- no NVSS transients brighter than 120\,mJy --- because when looking for transients in NVSS not present in ATATS, the comparison to the more sensitive master mosaic is the most appropriate.

If we consider sources which may have been transients in ATATS, but were not seen in NVSS, in \paperi\ we detected no ATATS transients brighter than 40\,mJy. The conservative single-epoch completeness limit derived in this paper, however, allows us to place more stringent limits on short-duration ATATS transients. A transient detected in a single ATATS epoch at 232\,mJy would have flux density $232 / 11 = 21$\,mJy in the master mosaic, assuming, that is, that its measured flux density in the master mosaic simply averaged down in such a way. Of course, a source which remained steady over several months at $\gtrsim 40$\,mJy would be easily detected in the master mosaic but likely missed in the individual ATATS epochs. But for bright sources with durations between a minute (the integration time per epoch, per pointing) to a few days (the cadence of the epochs), examination of the individual epoch catalogs provides the strongest constraint (about a factor two better than the flux density limit from \paperi). 

Following \citet{bower} and \citet{pigss}, the two-epoch rate for a survey with $N_e$ epochs that cover an area $A$ to sensitivity $S$ is 

\begin{equation}
R(>S)=\frac{N_t}{(N_e-1) A(>S)}
\label{eqn:snapshot}
\end{equation}

where $N_t$ is the number of transients detected.  Here, $A(>S)$ refers to 
the solid angle over which a source of flux density $S$ or greater can 
be detected.  Where no transient is 
detected, the $2\sigma$ limit is $N_t \approx 3$.

Although the completeness limits in Table~\ref{tab:epochs} are in some cases as low as $\sim 100$\,mJy, we choose the most conservative value, 232\,mJy, because a source with a constant flux density of 110\,mJy might appear to be a transient if it only showed up in a few epochs. We have 11 independent epochs, with good coverage of 564~\sqdegs\ of sky, yielding a $2\sigma$ upper limit to the transient rate (for transients with peak flux density $\gtrsim 232$\,mJy and timescales of minutes to days) of $5.3 \times 10^{-4}$~deg$^{-2}$.

\begin{figure}
\centering
\includegraphics[width=\linewidth,draft=false]{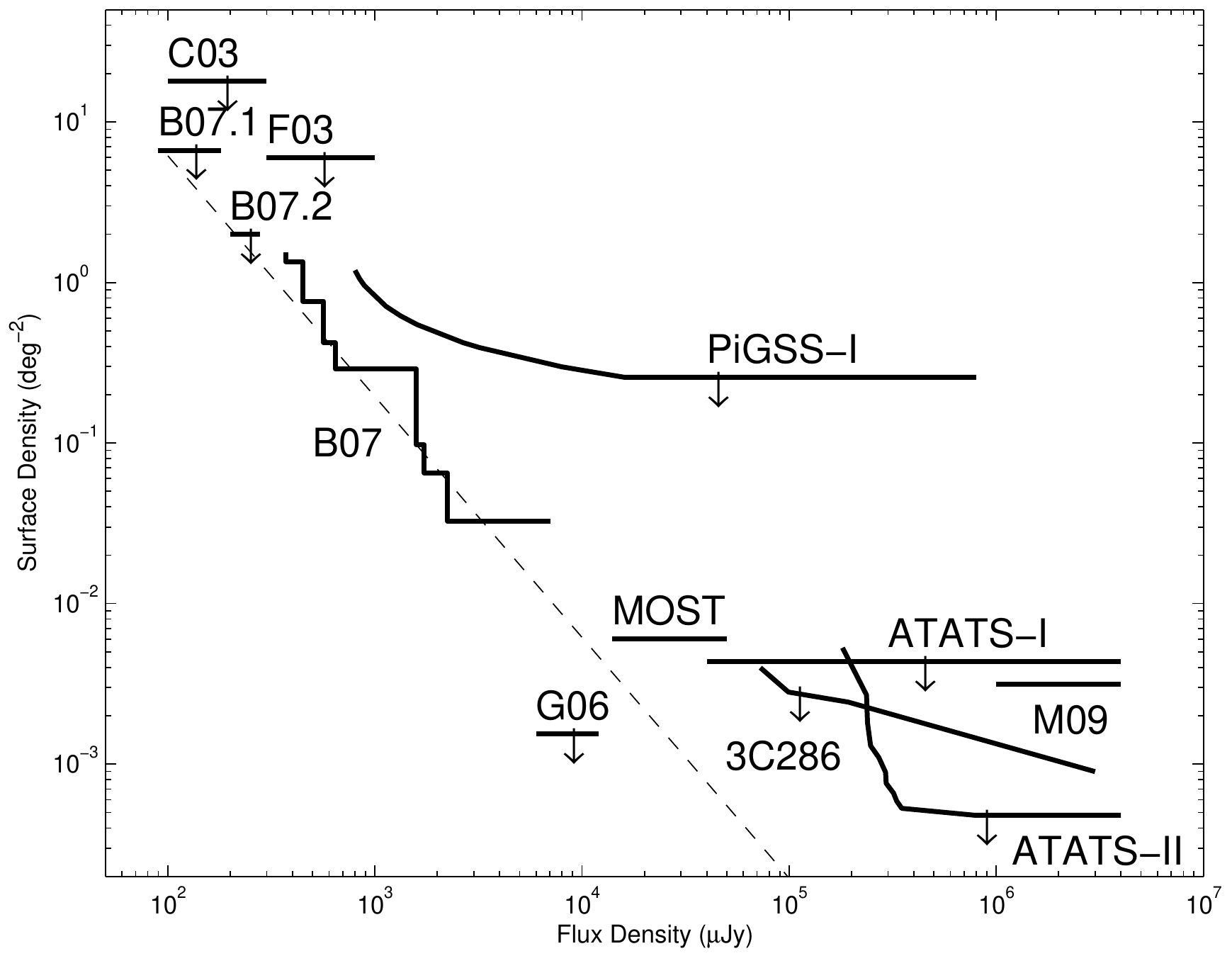}
\caption{\label{fig:rate}
Cumulative two-epoch source density for radio transients as a
function of flux density, based on Fig.~9 of \citet{bower} and Fig.~20 of \paperi. The solid black line labeled B07 shows the rate measured for transients with characteristic timescales $< 7$~days from \citet{bower}. The dashed line shows an $S^{-1.5}$ curve fixed at the B07 rate at 1.5\,mJy. The arrows show $2 \sigma$ upper limits for
transients from \citeauthor{bower} with a 1\,yr timescale (labeled B07.1) and a 2 month timescale (labeled B07.2), and for
transients from the comparison of the 1.4\,GHz NVSS and FIRST surveys by \citet{galyam}, labeled G06; from the \citet{carilli:03} survey (labeled C03); from the \citet{frail:03} survey (labeled F03); from the PiGSS-I survey \citep{pigss}; and from a VLA survey of the 3C\,286 field \citep{3c286}. Also shown are rates reported from the MOST archive by \citet{bannister:10} and from the \citet{matsumura:09} survey (labeled M09). The upper limits from \paperi\ (labeled ATATS1) are only marginally consistent with the \citeauthor{matsumura:09} detections. The upper limits from this paper (labeled ATATS2) appear to rule out an astronomical origin for the \citeauthor{matsumura:09} transients. 
}
\end{figure}

However, this rate relies on the somewhat subjective classification of the 22 candidates not within 4\degr\ of 3C\,286 or 3C\,295 as unlikely to be real astronomical events. Although we are quite confident that this is indeed the case, we can compute slightly weaker constraints on transient rates in a more robust manner by not relying on the classification of any candidates by eye. In Table~\ref{tab:epochs}, we tabulate the brightest transient candidate seen in each epoch (but in no other epochs, including NVSS) over the 564~\sqdegs\ of good coverage. No epoch has a transient brighter than 792\,mJy. Indeed, the one 791.5\,mJy candidate seen (in epoch ATA7) is only 4.1\degr\ from 3C\,295 and so is likely to be a sidelobe source and not an astronomical event. But whether or not this event is astronomical, we can still use Equation~\ref{eqn:snapshot} to compute \label{sec:robust} an upper limit for the two-epoch rate of $5.3 \times 10^{-4}$ for transients of 792\,mJy or brighter. Similarly, the 10 epochs with no transient candidate brighter than 350\,mJy, and the 9 epochs with no transient candidate brighter than 330\,mJy, can be used to compute two-epoch rate upper limits over decreasing effective areas (due to the decreasing numbers of epochs) but to fainter flux densities. We tabulate these rates as a function of number of epochs in Table~\ref{tab:complete}, and plot them in Fig.~\ref{fig:rate}.

\begin{deluxetable}{lll}
\tablewidth{0pt}
\tabletypesize{\scriptsize}
\tablecaption{\label{tab:complete} Two-epoch transient rate limits as a function of flux limit}
\tablehead {
\colhead{Flux limit (mJy)} & 
\colhead{Number of epochs (including NVSS)} & 
\colhead{Two-epoch rate $2\sigma$ upper limit (deg$^{-2}$)}
}
\startdata
792 & 12 & $4.8 \times 10^{-4}$ \\
350 & 11 & $5.3 \times 10^{-4}$ \\
330 & 10 & $5.9 \times 10^{-4}$ \\
319 & 9 & $6.6 \times 10^{-4}$ \\
294 & 8 & $7.6 \times 10^{-4}$ \\
292 & 7 & $8.9 \times 10^{-4}$ \\
271 & 6 & $1.1 \times 10^{-3}$ \\
247 & 5 & $1.3 \times 10^{-3}$ \\
239 & 4 & $1.8 \times 10^{-3}$ \\
236 & 3 & $2.7 \times 10^{-3}$ \\
181 & 2 & $5.3 \times 10^{-3}$ \\
\enddata
\end{deluxetable}

Fig.~\ref{fig:rate} shows the constraints on transient rates from some of the surveys discussed in \paperi, compared to the constraint from the comparison of ATATS and NVSS from \paperi, and now with the addition of the two-epoch rates from Table~\ref{tab:complete}. We see that examination of the individual ATATS epochs strongly rules out the 9 transients brighter than 1\,Jy reported by \citet{matsumura:09}. The characteristic timescale of the transients reported by \citeauthor{matsumura:09} is minutes to days, so in contrast to \paperi\ we are probing the same timescale here. \citeauthor{matsumura:09} report an areal density for their transients of $\sim 8.7 \times 10^{-7}$\,arcmin$^{-2}$ (although there is some uncertainty in how this rate was derived --- see \citealt{3c286}), which is $3.1 \times 10^{-3}$\,deg$^{-2}$. If their transients are truly astronomical events, we would expect to see $17 \pm 4$ events brighter than 1\,Jy in ATATS. Even if the single epoch ATATS images underestimated the flux densities of these sources by a factor 2 (see Fig.~\ref{fig:flux0111}), such events would show up clearly in our data. If the flux densities of the \citeauthor{matsumura:09} candidates follow a typical power law distribution, we would expect to see approximately 80 events brighter than 350\,mJy. The \citeauthor{matsumura:09} candidates are found at both high and low Galactic latitude, so if some of their candidates are Galactic and some extragalactic, the discrepancy in the rates might be reduced somewhat, but presumably not enough to make them agree.

\section{Summary} 

We examined images and catalogs from the 11-epoch ATATS survey, which was taken during ATA commissioning. ATATS explores the challenges of multi-epoch transient and variable source surveys in the domain of dynamic range limits and changing snapshot \uv\ coverage. To distinguish true astronomical transients from spurious sources requires that we restrict analysis to relatively bright sources, or to sources confirmed to be astronomical due to their detection in multiple epochs. 

Detection of transient candidates in at least two epochs is a good filter to remove false positives due to calibration problems, and a good first step for any many-epoch search for transients with characteristic timescales similar to the cadence between epochs. Confirmation of transient candidates could also be obtained by simultaneous monitoring by another telescope, either a radio telescope at another site pointed at the same field (which in principal could allow robust detections of transients below the formal completeness limit of either survey alone), or follow-up at another wavelength. The development of reduction and analysis pipelines that operate in close to real time will enable rapid follow-up of candidates to confirm whether or not they are real.

In designing future surveys, careful thought should be given to ensuring sufficient \uv\ coverage to enable good calibration (an issue for ATATS due to ongoing telescope commissioning at the time of the observations, and the choice of a single snapshot per field, per epoch, rather than multiple visits). Good calibration is essential to avoid samples of transient candidates being overwhelmed by false positives --- while transient candidates may be a small fraction of the total number of sources detected, they can still potentially swamp the true transient rate.

It is also important, when comparing transient rates from different surveys, to be aware of differing survey flux density limits and the range of timescales to which surveys are sensitive. A source which appears as a variable in one survey may appear as a transient in a survey with a slightly brighter flux density limit. Transient surveys are sensitive to sources with a range of timescales, from the integration time of the observations, through the total time spent observing a field per epoch, and the cadence of the observations, to the total time from the first to the last epoch, and once again, sources which appear as variables or transients on some timescales may not appear so on others.

Despite such caveats, analysis of the ATATS survey, a pilot for the current generation of ATA surveys such as PiGSS \citep{pigss}, enables us to place new constraints on transient rates at 1.4\,GHz flux densities of brighter than a few hundred mJy. No sources brighter than 350\,mJy present in a single ATATS epoch but not present in NVSS (within a 75\arcsec\ match radius) were seen in 10 of the ATATS epochs. This enables us to place a $2\sigma$ upper limit on transients with timescales of $\sim 1$ day that are brighter than 350\,mJy of $6 \times 10^{-4}$ deg$^{-2}$, strongly ruling out an astronomical origin for the $\sim 1$\,Jy transients reported by \citet{matsumura:09}, based on their reported rate.

Additionally, inspection of transient candidates appearing in one or more ATATS epochs with flux densities brighter than 232\,mJy, but absent from NVSS, suggests that none of these candidates are astronomical events.

\acknowledgments

The authors would like to acknowledge the generous support of the Paul
 G. Allen Family
 Foundation, which has provided major support for the design, construction,
 and operation of
 the ATA. Contributions from Nathan Myhrvold, Xilinx Corporation, Sun
 Microsystems,
 and other private donors have been instrumental in supporting the ATA.
 The ATA has been
 supported by contributions from the US Naval Observatory in addition
 to National Science
 Foundation grants AST-050690, AST-0838268 and AST-0909245.
 
 We are grateful for the support of the entire ATA team: 
 Rob Ackermann,
 Shannon Atkinson,
 Peter Backus,
 Billy Barott,
 Amber Bauermeister,
 Samantha Blair,
 Leo Blitz,
 Douglas Bock,
 Tucker Bradford,
 Calvin Cheng,
 Chris Cork,
 Mike Davis,
 Dave DeBoer,
 Matt Dexter,
 Gregory Desvignes,
 John Dreher,
 Greg Engargiola,
 Ed Fields,
 Matt Fleming,
 James R. Forster,
 Colby Gutierrez-Kraybill,
 Gerry Harp,
 Carl Heiles,
 Tamara Helfer,
 Chat Hull,
 Jane Jordan,
 Susanne Jorgensen,
 Tom Kilsdonk,
 Joeri van Leeuwen,
 John Lugten,
 Dave MacMahon,
 Peter McMahon,
 Oren Milgrome,
 Tom Pierson,
 Karen Randall,
 John Ross,
 Seth Shostak,
 Andrew Siemion,
 Ken Smolek,
 Jill Tarter,
 Douglas Thornton,
 Lynn Urry,
 Artyom Vitouchkine,
 Niklas Wadefalk,
 Sandy Weinreb,
 Jack Welch, and
 Dan Werthimer.
 
 We thank the anonymous referee for helpful comments and suggestions.
 
 We dedicate this paper to the memory of our friend and colleague Don Backer, whose enthusiasm for the ATA, its capabilities, and its potential, continue to inspire us.

\end{document}